\documentclass[aps,twocolumn,showpacs,preprintnumbers,floatfix, prb]{revtex4-2}

\usepackage{graphicx}
\usepackage{dcolumn}
\usepackage{bm}
\usepackage{setspace}

\DeclareMathAlphabet{\mathpzc}{OT1}{pzc}{m}{it}

\usepackage[caption=false]{subfig}
\usepackage{amssymb}
\usepackage{amsmath}
\usepackage{graphicx,bm}
\usepackage{verbatim}
\usepackage{xcolor, soul}
\sethlcolor{yellow}
\usepackage{amsfonts}
\usepackage{natbib}

\usepackage{float}

\begin{document}

\title{Optimal reduction of an epidemic outbreak size via temporary quarantine}

\author{Eyal Atias}
 \email{eyal.atias1@mail.huji.ac.il}

\author{Michael Assaf}\email{michael.assaf@mail.huji.ac.il}
\affiliation{Racah Institue of Phyiscs, Hebrew University of Jerusalem, Jerusalem 91904, Israel}

\begin{abstract}
    Understanding the dynamics of an epidemic spread is crucial for effective control measures. During the COVID-19 pandemic, quarantines were implemented to minimize infections while mitigating social and economic impacts, raising the question of how to maximize quarantine efficiency. Previous research on periodic quarantines using the susceptible-infected-recovered (SIR) and similar models identified optimal duration for periodic quarantines. However, the question of the optimal \textit{initiation} time for a single quarantine remains unanswered.
Here, we use the SIR model in order to determine the optimal quarantine initiation time, by computing the optimal susceptible fraction at the onset of the quarantine, which minimizes the total outbreak size. Our analysis extends from a well-mixed scenario to strongly-heterogeneous social networks.
We show that the optimal quarantine initiation time is closely related to the so-called ``herd immunity" threshold, occurring at the onset of epidemic decline. Importantly,  providing a methodology for identifying the optimal quarantine initiation time across different network structures, entails significant implications for epidemic control.
\end{abstract}
\maketitle

\section{\label{sec:Intro}Introduction}
Modeling and predicting the spread of pandemics has long intrigued researchers, leading to the development of various deterministic as well as stochastic epidemiological models~\cite{anderson1991infectious,kermack1927contribution,mollison1995epidemic,daley1999epidemic,hethcote2000mathematics,keeling2005networks,allen2008introduction,andersson2012stochastic}. One of the most fundamental models is the susceptible-infected-susceptible (SIS) model~\cite{hethcote2000mathematics}, which divides the population into two compartments: susceptibles (individuals who can get infected) and infected (individuals who can transmit the infection to susceptibles and eventually become susceptible again). This model is suitable for diseases such as influenza, or the common cold, with relatively short-term immunity such that individuals can be reinfected after being recovered. 

A more realistic model which is widely used to characterize the spread of epidemics is the susceptible-infected-recovered (SIR) model, which complements the SIS model with a third \textit{recovered} compartment~\cite{andersson2012stochastic}. Here, the underlying assumption is long-term immunity such that recovered individuals do not get reinfected immediately. 
Naturally, the SIR can be generalized, e.g., by including the possibility
of reinfection, or adding an \textit{exposed} phase prior to infection, to account for an incubation period of the disease~\cite{andersson2012stochastic}. The SIR model and its 
generalizations are amply used to describe the dynamics of a wide variety of diseases such as measles, chickenpox, rubella, smallpox, polio, and severe acute
respiratory syndrome (SARS), including the recent COVID-19 pandemic~\cite{yang2021rational,hindes2022outbreak}.

One of the most important factors that determine the disease dynamics is the topology of the population network, namely, the structure of the social contacts within the population. The simplest form of topology is the so-called ``well-mixed" setting, where each individual is equally connected to all other individuals~\cite{anderson1991infectious,hethcote2000mathematics,pastor2015epidemic,saeedian2017memory,bohner2019exact}. Indeed, this setting holds, e.g., when the population is confined to a relatively small area, and each individual can interact with all others. 
Yet, a more realistic scenario includes a nontrivial population network, where each individual is represented by a node, and has a certain \textit{degree}, which determines the number of contacts, or edges, through which infection can spread.  Here, the network is often characterized by a degree distribution, providing information about the likelihood of having a specific number of connections. 
For generic degree distributions these networks are called \textit{heterogeneous}, providing a clear departure
from classic compartmental models, such that key quantities including epidemic and herd-immunity thresholds,  expected outbreak size, and disease lifetime, may dramatically vary, see e.g.,~\cite{keeling1999effects,barabasi1999emergence,pastor2001epidemic,moore2000epidemics,newman2002assortative,hindes2019degree}. Notable examples for such population networks are homogeneous networks, in which each node has an equal degree, Erd\"os-R\'enyi networks~\cite{erdos1959random} with a Poisson degree distribution (for large networks), gamma-distributed~\cite{Gammathom1958note} and power-law~\cite{PowerLawadamic2001search,julicher2020} networks. Notably, the well-mixed setting is a particular example of a homogeneous network where the degree of each node equals the population size.

While the well-mixed scenario has been extensively analyzed, and both  the expected outbreak size, as well as its distribution have been rigorously found~\cite{anderson1991infectious,hethcote2000mathematics,pastor2015epidemic,saeedian2017memory,bohner2019exact,ball1986unified,ball1993final,keeling2011modeling,house2013big,allen2017primer,miller2019distribution,hindes2022outbreak,hindes2023outbreak}, dealing with the SIR model on heterogeneous networks is more intricate.
Here, most works dealing with generic networks have focused on the  deterministic (or mean-field) dynamics, while keeping the infection and recovery rates constant~\cite{newman2002spread,meyers2005network,kenah2007second,volz2008sir,miller2011note}. 
Recently, a generalized SIR model with time-fluctuating infection and recovery rates has been studied, in a well-mixed setting~\cite{hindes2023outbreak}. Yet, a systematic analysis of the SIR model on heterogeneous population networks with explicitly time-dependent rates, has not been carried out so far. 

In this work we focus on the SIR model on generic population networks, and study how the mean-field dynamics, and in particular, the expected outbreak size, are affected when a temporary quarantine is introduced during the epidemic wave. 
The quarantine is implemented by abruptly decreasing the infection rate by a given factor at some given time and for a prescribed duration, due to intervention measures such as increased seclusion~\cite{barzelcovid2021}. Subsequently, the quarantine is lifted and the infection rate goes back to its pre-quarantine value. 

The primary aim of implementing quarantine measures is to mitigate the spread of infection and thus control the epidemic. 
Here, a key question is how to make the quarantine most efficient, given that its duration and magnitude are fixed. Recently there have been some works dealing with quarantines and ways to optimize them. In Ref.~\cite{yan2007optimal} the authors have dealt with a generalized SIR model and studied how the overall cost of a quarantine can be minimized. In~\cite{barzelcovid2021,hindes2021optimal} periodic quarantines were studied: in Ref.~\cite{hindes2021optimal} the optimal quarantine duration  was computed using a generalized SIR model, whereas in Ref.~\cite{barzelcovid2021} the authors have studied an alternating quarantine (each time on half of the population) and how it can be optimally synchronized with the disease lifetime.

We here study a different angle and ask the following question: given a quarantine of prescribed duration and magnitude, is there an optimal \textit{initiation} time of the quarantine, such that the expected final outbreak size is minimized? Remarkably, our results reveal that indeed there exists an optimal quarantine onset, which gives rise to a minimization of the overall expected number of infected  during the epidemic wave. 
Notably, since the fraction of susceptibles is a monotone decreasing function of time, the optimal initiation time of the quarantine can be converted into an optimal susceptible fraction. Once this optimal fraction is reached, one can strategically time the initiation of quarantine measures, holding significant implications for managing future pandemics.

This rest of the paper is organized as follows. Section~\ref{sec:theory} introduces the theoretical model and provides a methodology for determining the optimal moment to initiate quarantine, for a variety of network topologies. Section~\ref{sec:results}  details the results for the different networks and our numerical algorithm used to validate the theoretical findings. Finally, 
Sec.~\ref{sec:discussion} presents our conclusions, and offers an explanation and interpretation of the results.

\section{\label{sec:theory}Theoretical formulation}\vspace{-2mm}
\subsection{Well-mixed setting}
In the SIR model, the  population is comprised of three compartments: $\mathcal{S}$, $\mathcal{I}$ and $\mathcal{R}$,  denoting the numbers of susceptibles, infected and recovered in the population, respectively. In the absence of demography, the total population size is conserved: $\mathcal{S} +\mathcal{I} + \mathcal{R} = N$, where $N$ is the total size of the population network, and we assume throughout this work that $N\gg 1$. In a well-mixed setting, the probability per unit time that the number of susceptibles decreases by one and the number of infected increases by one is $\beta \mathcal{S} \mathcal{I}/N$, where $\beta$ is the infection rate per individual. Similarly, the probability per unit time that the number of infected decreases by one is $\gamma \mathcal{I}$, where $\gamma$ is the recovery rate per individual. Combining both processes results in a discrete-state system with the following stochastic reactions:
\begin{eqnarray}
    &&(\mathcal{S},\mathcal{I}) \to (\mathcal{S}-1,\mathcal{I}+1) \text{    with rate    }  \beta \mathcal{S} \mathcal{I}/N \nonumber\\
    &&(\mathcal{I},\mathcal{R}) \to (\mathcal{I}-1,\mathcal{R}+1) \text{    with rate    } \gamma \mathcal{I}.
\end{eqnarray}
For simplicity, we introduce the fractions of susceptibles $
S = \mathcal{S}/N$, infected $I = \mathcal{I}/N$, and recovered $R = \mathcal{R}/N$. Denoting the basic reproduction number $R_0 = \beta/\gamma$, rescaling time $t \to \gamma t$, and assuming a well-mixed setting of a fully-connected population of $N\gg 1$ individuals, the deterministic rate equations read:
\begin{equation}
    \label{eqn:SIR}
    \dot{S} = -\mathfrak{R}(t) SI ,\quad \dot{I} = \mathfrak{R}(t) SI - I ,\quad \dot{R} = I.
\end{equation}
These equations require an initial condition: we assume an initial fractions of $I_0$ infected, and $S(t=0) = 1-I_0$ susceptibles. Our aim is to determine the \textit{optimal} susceptible fraction, $S_\text{opt}$, at which the quarantine is initiated, such that the final outbreak size is minimized. 

As the epidemic progresses, a quarantine is initiated at time $t_0>0$, when the fraction of susceptibles right \textit{before} the quarantine is denoted by $S_b\equiv S(t=t_0)$. At this point, the infection rate drops from $\beta$ to $\xi \beta$, with $0\leq\xi<1$, for a prescribed duration $\Delta t$. At $t=t_0+\Delta t$ the infection rate returns to its pre-quarantine value, until the epidemic wave is over.
At the deterministic level, the dynamics satisfy Eq.~(\ref{eqn:SIR}), with an explicitly time-dependent basic reproduction number $\mathfrak{R}$:
\begin{equation}\label{frakr}
   \mathfrak{R}(t) =
    \begin{cases}
        R_0 & t<t_0\;\; \text{or} \;\; t>t_0 + \Delta t \\
        R_0 \xi & t_0<t<t_0+ \Delta t, 
    \end{cases}
\end{equation}
where $R_0$ is the basic reproduction number in the pre- and post-quarantine stages.

We now determine the final outbreak fraction of the epidemic, denoted as $R_\infty = R(t \to \infty)$, as a function of $I_0$, $\xi$, $S_b$, $R_0$, and $\Delta t$. 
Following Ref.~\cite{julicher2020}, we define a new time scale $\tau$ such that $\dot{\tau} = \mathfrak{R}I$. This readily yields:
\begin{equation}\label{stau}
    S(\tau) = (1-I_0)e^{-\tau}.
\end{equation}
While $S(\tau)$ takes a rather simple form, $I(\tau)$ has to be determined in three separate regimes: before, during, and after the quarantine, since $\mathfrak{R} = \mathfrak{R}(t)$. Demanding continuity of $I(\tau)$, after some algebra we find:
\begin{eqnarray}
\label{eqn:Iwell}
 && I(\tau) = (I_0 - 1) e^{-\tau} + f(\tau), \\
 && f(\tau) = 
    \begin{cases}
          -\frac{\tau}{ R_0} + 1 & \tau < \tau_0, \nonumber\\ 
  -\frac{\tau}{ R_0 \xi} + 1 + \frac{\tau_0}{ R_0} \left(\frac{1}{\xi} - 1\right) &  \tau_0 < \tau < \tau_0 + \Delta \tau, \nonumber\\ 
  -\frac{\tau}{ R_0} + 1 + \frac{\Delta \tau}{ R_0} \left(1 - \frac{1}{\xi}\right) &  \tau > \tau_0 + \Delta \tau.     
    \end{cases}
\end{eqnarray}
Here $\tau_0$ is found from $S(\tau)$ and reads $\tau_0 = \ln[(1-I_0)/S_b]$. In addition, $\Delta \tau$ can be  determined using the equation for $\dot{\tau} = \mathfrak{R}I=\xi  R_0 I$ during the quarantine, leading to:
\begin{equation}
\label{eqn:dteq}
    \hspace{-2mm}\int_{\tau_0}^{\tau_{0}+\Delta\tau}\!\!\!\!\!\!\! \frac{d\tau'}{1 + \frac{\tau_0}{ R_0} \left(\xi^{-1} \!-\! 1\right) - (1 \!-\! I_0) e^{-\tau'} \!-\! \frac{\tau'}{ R_0 \xi}} =  R_0 \xi \Delta t.
\end{equation}
This allows to compute $R_{\infty}$ by using the fact that $\dot{R}=I$ and $\dot{\tau} =  \mathfrak{R}(t) \dot{R}$, which yields $R(\tau) = \tau/\mathfrak{R}(t)+C$, where $C$ is a constant. Demanding continuity of $R(\tau)$, we find:
\begin{equation}
    R(\tau) = 
    \begin{cases}
    \frac{1}{ R_0} \tau & \tau < \tau_0, \\
    \frac{1}{ R_0 \xi} \tau + \frac{\tau_0}{ R_0} \left(1-\frac{1}{\xi}\right) & \tau_0 < \tau < \tau_0 + \Delta\tau, \\
    \frac{1}{ R_0} \tau + \frac{\Delta\tau}{ R_0} \left(\frac{1}{\xi} - 1\right) & \tau > \tau_0 + \Delta\tau.
    \end{cases}
\end{equation}
Using the last regime, we obtain $\tau_\infty \!=\!  R_0 R_\infty \!+\! (1\!-\!\xi^{-1}) \Delta\tau$, 
which, with $S_\infty = (1-I_0)e^{-\tau_\infty}$ and $R_\infty = 1-S_\infty$, yields a closed-form equation for $R_\infty$. The solution satisfies:
\begin{equation}
\label{eqn:Rinfwell}
    R_\infty = 1 + \frac{1}{ R_0} W_0 \left[ - R_0 e^{- R_0} (1-I_0) e^{(\xi^{-1}-1)\Delta\tau} \right],
\end{equation}
where $W_0(x)$ is the main branch of Lambert's $W$ function, and $\Delta \tau$ is given by Eq.~(\ref{eqn:dteq}), with $\tau_0$ given right below Eq.~(\ref{eqn:Iwell}). Note that,  in the absence of a quarantine (i.e., $\xi=1$), we arrive at the known results~\cite{julicher2020}:
\begin{eqnarray}
    \label{eqn:knownJullicher}
    &&I(\tau) = (I_0 - 1) e^{-\tau}-\frac{\tau}{ R_0} + 1 ,\quad R(\tau) = \frac{1}{ R_0} \tau \,\nonumber\\
     &&R_\infty = 1 + \frac{1}{ R_0} W_0 \left[ - R_0 e^{- R_0} (1-I_0) \right].
\end{eqnarray}

\begin{figure}[h!]  
    \centering
    \includegraphics[width=.93\linewidth]{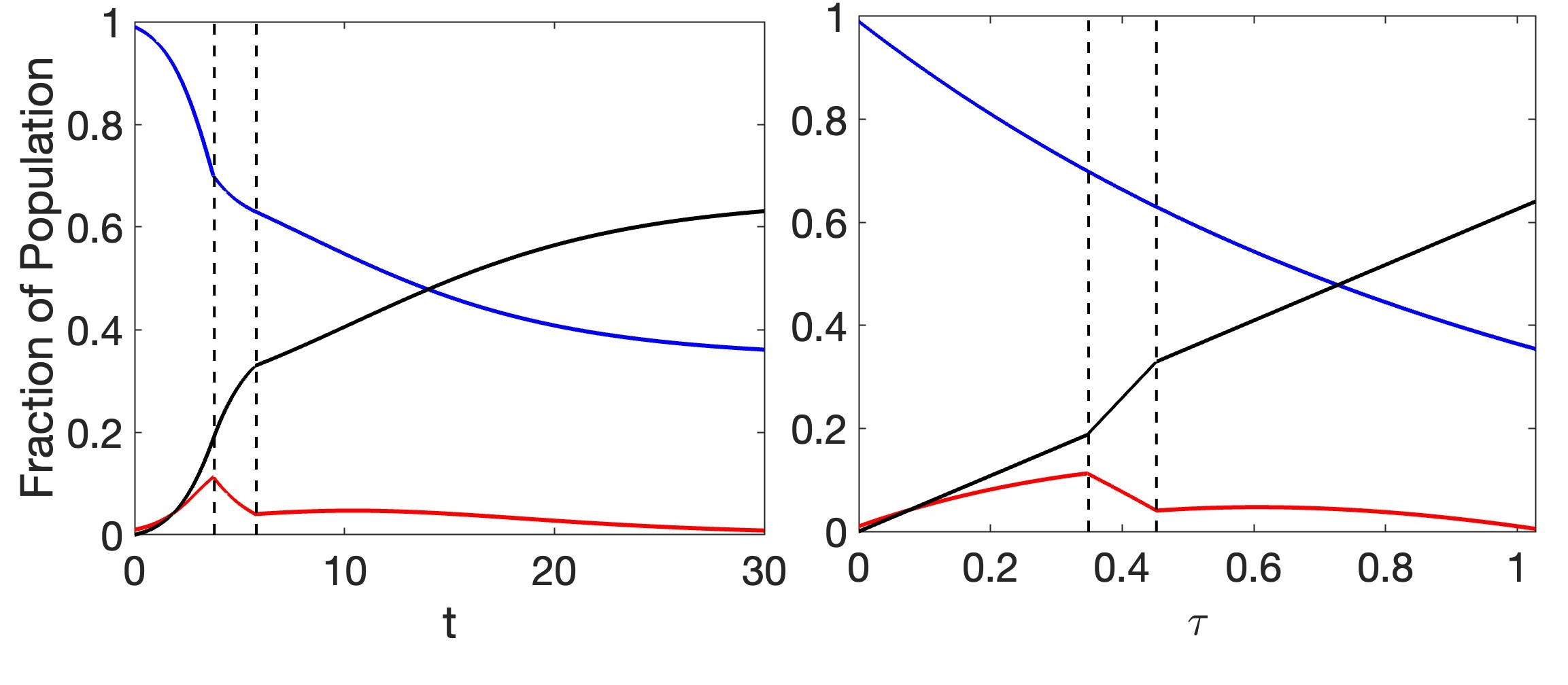} 
\vspace{-6mm}
    \caption{SIR dynamics in the presence of a temporary quarantine. Shown are fractions of susceptibles (blue lines), infected (red lines) and recovered (black lines) versus time (left) and $\tau$ (right). Here $R_0 = 1.85$, $\xi = 0.4$, $S_b = 0.7$, and $ \Delta t = 2$.
    Dashed vertical lines mark the quarantine boundaries.} \label{fig1}
\end{figure}

In Fig.~\ref{fig1} we illustrate an example of an epidemic outbreak. One can see the fractions $S$, $I$ and $R$ versus time (left) and $\tau$ (right), where the latter depends on time nontrivially. Using the fact that $\dot{\tau} =  \mathfrak{R}(t) I$, we find:
\begin{equation}
    \int_{0}^{\tau} \frac{d\tau'}{I(\tau')} = \int_{0}^{t} \mathfrak{R}(t)dt, 
\end{equation}
where  $I(\tau)$ and $\mathfrak{R}(t)$ are given by Eqs.~(\ref{frakr}) and (\ref{eqn:Iwell}).
In this figure, a quarantine is introduced when the fraction of susceptibles is $70\%$, i.e. $S_b=0.7$, such that the infection rate during the quarantine drops to $40\%$ its pre-quarantine value. As expected, a discontinuity in the derivatives of $S(t)$ and $I(t)$ is displayed, at the start and end of the quarantine. Note that, in the right panel of Fig.~\ref{fig1}, the fractions are plotted versus $\tau(t)$, which is proportional to $R(t)$, hence the linear piecewise behavior of $R(t)$.
In contrast, $S(\tau)$ remains smooth as function of $\tau$ in all three regimes [see Eq.~(\ref{stau})].

To find the optimal susceptible fraction at the quarantine onset, $S_{\text{opt}}$, we differentiate Eq.~(\ref{eqn:Rinfwell}) with respect to $S_b$. Since only $\Delta \tau$ depends on $S_b$ in Eq.~(\ref{eqn:Rinfwell}) via Eq.~(\ref{eqn:dteq}) (as $\tau_0$ depends on $S_b$),  $dR_{\infty}/dS_b$ reads:
\begin{equation}\label{diffRinfty}
        \frac{d R_\infty}{d S_b} =  \frac{1}{ R_0}\frac{ W_0 \left( x \right)}{1+ W_0 \left( x\right)}  \frac{d\Delta \tau}{dS_b},
\end{equation}
where $x =  - R_0 e^{- R_0} (1-I_0) e^{(\xi^{-1}-1)\Delta\tau}$. 
Equating~(\ref{diffRinfty}) to zero allows to determine $S_{\text{opt}}$ that minimizes $R_\infty$. To this end, as $W_0(x\neq 0)\neq 0$, it is sufficient to demand that $\Delta \tau'(S_b) = 0$, where $\Delta \tau(S_b)$ is given by Eq.~(\ref{eqn:dteq}). 

In general, for an arbitrary quarantine duration, computing this derivative can only be done numerically.
Yet, for $\Delta t\ll 1$, we can approximate the integral in Eq.~(\ref{eqn:dteq}) by Taylor-expanding the result at $\Delta\tau\ll 1$~\footnote{Going back to physical time units, this requirement means that that the quarantine duration is much shorter than the typical time for recovery.}. Since the denominator of~(\ref{eqn:dteq}) equals $I(\tau')$, for $\Delta\tau\ll 1$ we have:
\begin{equation}
    \int_{\tau_0}^{\tau_0+\Delta\tau}\frac{d\tau'}{I(\tau')}\simeq \frac{\Delta\tau}{I(\tau_0)}-\frac{(\Delta\tau)^2}{2}\frac{I'(\tau_0)}{I(\tau_0)^2}.
\end{equation}
Plugging this approximation into Eq.~(\ref{eqn:dteq}) allows us to find $\Delta\tau$ as function of $\Delta t$ and $S_b$, and subsequently find $d\Delta \tau/dS_b$. Equating this derivative to zero, and solving peturbatively with respect to $\Delta t\ll 1$, we are able to find the leading- and subleading-order terms in $S_{\text{opt}}$ order by order in $\Delta t\ll1$~\cite{2008perturbation}. The result is
\begin{equation}
\label{eqn:ScOptSmallDtWell}
    S_{\text{opt}} \simeq \frac{1}{R_0}+\left( \frac{R_0-\ln{R_0}-1}{2 R_0}\right) \xi \Delta t.
\end{equation}
Importantly, in the limit $\Delta t\to 0$, the optimal susceptible fraction at the quarantine onset coincides with the herd immunity (HDI) threshold  $S_{\text{hdi}}$, which equals $1/R_0$ for fully-connected networks in the absence of quarantine measures. This threshold represents the point at which the infected population reaches its peak and starts declining. Our result indicates that the longer the quarantine duration is, the earlier it has to be initiated, before the HDI threshold is reached, in order to minimize $R_{\infty}$.

\subsection{Heterogeneous Networks}
We now determine $S_{\text{opt}}$ for heterogeneous population networks. Here, unlike a well-mixed population, each individual is connected according to a prescribed degree distribution $p_k$, denoting the probability for a node in the network to have $k$ connections, such that $\sum_k p_k = 1$. 

When dealing with population networks it is convenient to define the infection rate $\beta$ per contact (rather than per individual), whereas $\gamma$ still denotes the recovery rate per individual, and time is measured in units of $\gamma^{-1}$.
Denoting $k_0=\sum_k kp_k$ and $\sigma^2=\sum_k k^2p_k-k_0^2$ as the distribution's mean and variance, the basic reproduction number $R_0$ (in the absence of a quarantine) satisfies:
\begin{equation}
\label{eqn:R0}
    R_0 = \beta/\beta_c, \quad \beta_c = k_0/(k_0^2+\sigma^2-2).
\end{equation}
Here, $\beta_c$ is the critical infection rate below which the epidemic dies out instantly~\cite{pastor2015epidemic, morita2022basic,leibenzon2024heterogeneity}.
For simplicity we will focus on networks with $k_0\gg 1$ such that $\beta_c\simeq k_0/(k_0^2+\sigma^2)$. Notably, when $\sigma=0$ and $k_0=N$, $\beta_c=1/N$ and $R_0=N\beta$, as expected in the well-mixed case.

\begin{figure}[H]
    \centering
    \includegraphics[width=.9\linewidth]{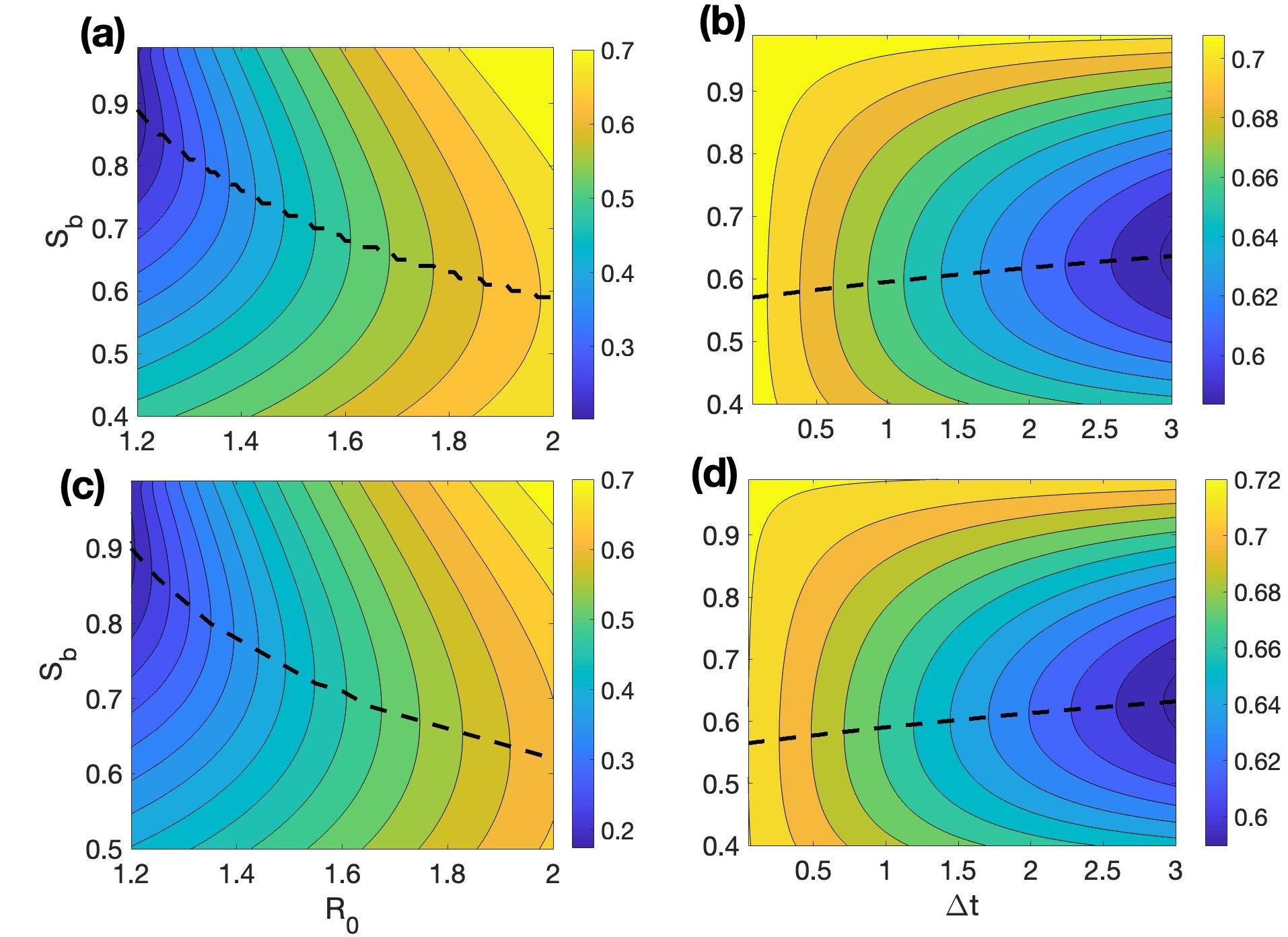}
    \vspace{-5mm}
    \caption{Heatmaps of $R_{\infty}$ for two different population networks. Shown is $R_{\infty}$ versus  $S_b$ and $R_0$ (a,c) and $S_b$ and $\Delta t$ (b,d). Upper (a,b) and lower (c,d) panels show results for a Poisson and gamma network, respectively. In all panels $k_0 = 50$ and $\xi = 0.5$. In (a,c) $\Delta t = 2$, while in (b,d) $R_0=1.85$. In (c,d) the gamma network has a COV of $\epsilon = 0.27$, see text. The black dashed lines denote $S_{\text{opt}}$.}
    \label{fig2}
\end{figure}

To illustrate the dependence of $R_{\infty}$ on $S_b$ and to show the existence of an optimal $S_b$, we plot in Fig.~\ref{fig2} heatmaps depicting $R_{\infty}$ as function of $S_b$ and $R_0$ (a,c) and $\Delta t$ (b,d), for Poisson and gamma population networks, see below. The heatmaps clearly demonstrate the existence of $S_{\text{opt}}$ that minimizes the final outbreak fraction $R_{\infty}$, for a wide range of $R_0$ values and quarantine durations $\Delta t$. In fact, $S_{\text{opt}}$ grows (almost linearly) as function of $\Delta t$, see below, indicating that the quarantine should be initiated earlier as $\Delta t$ is increased. Conversely,   $S_{\text{opt}}$ decreases  as $R_0$ increases, suggesting that the quarantine should be delayed as the basic reproduction number grows.

To study the dependence of $R_{\infty}$ on $S_b$ we adopt the approach of Ref.~\cite{miller2011note}. We denote by $\theta(t)$  the fraction of edges that have not yet transmitted the infection up to time $t$. This definition is equivalent to the probability
that a node of degree $1$ is still susceptible at time $t$~\cite{miller2011note}.
Thus, the probability of an individual node with $k$ contacts to remain susceptible at time $t$ is given by $\theta^k$. As a
result, the fraction of susceptibles at time $t$ is given by
\begin{equation}
\label{eqn:St}
S(t) = \Psi(\theta(t)) = \sum_{k=1}^{\infty} p_k \theta(t)^k,
\end{equation}
where $\Psi(\theta)$ is the probability generating function of the degree distribution~\cite{miller2011note}. Our aim is to find $\theta(t)$, and subsequently $\theta_\infty$, such that
$R_\infty = 1 - \sum_{k=1}^{\infty} p_k \theta_\infty^k$.
Following the methodology in~\cite{miller2011note} the dynamics of $\theta(t)$ satisfies:
\begin{equation}
\label{eqn:dynamicsMiller}
\hspace{-3mm}\dot{\theta} = -\beta(t)\phi,\quad
\dot{\phi} = -\left[1+\beta(t)-(\beta(t)/k_0)\Psi''(\theta)\right]\phi,
\end{equation}
where $\phi$ denotes the fraction of edges that have not yet transmitted infection, and have an infected base node. Combining Eqs.~(\ref{eqn:dynamicsMiller}), the dynamics satisfy: 
\begin{equation}
    \label{eqn:milleq}
    \dot{\phi} = \left[1+\beta(t)^{-1}-(1/k_0)\Psi''(\theta)\right]\dot{\theta}.
\end{equation}
Given that the quarantine starts at time $t_0>0$ and ends at time $t_0+\Delta t$, Eq.~(\ref{eqn:milleq}) needs to be solved in three regimes: (i) pre-quarantine $0\leq t\leq t_0$ with an infection rate $\beta$; (ii) during the quarantine $t_0\leq t\leq t_0+\Delta t$ with an infection rate $\beta\xi$; and (iii) post-quarantine $t>t_0+\Delta t$ with an infection rate $\beta$. Naturally, the integration constants must be such that $\theta(t)$ is continuous at all times.

At this point we note that in addition to $\theta(t)$ being a continuous function of time, $\phi(t)$ must be continuous as well.  Suppose $\phi$  has a finite discontinuity at $t=t_0$ and at $t=t_0+\Delta t$. Since $\beta(t)$ has a finite discontinuity at those times as well, $\dot{\phi}$ will also have a finite discontinuity exactly at those times, as $\dot{\phi}$ is proportional to $\phi$, and $\Psi''(\theta)$ is a continuous function of $\theta$, see Eq.~(\ref{eqn:St}). Yet, if $\dot{\phi}$ has a finite discontinuity, its integral with respect to time, $\phi$, must be continuous at those times, in contradiction to our initial assumption. Hence, $\phi$ is continuous. 

Once we have established that both $\theta$ and $\phi$ are continuous functions of time, despite the discontinuity in the infection rate $\beta(t)$, we can integrate Eq.~(\ref{eqn:milleq}) between $t=0$ and $t=t_0$, between $t=t_0$ and $t=t_0+\Delta t$, and between $t=t_0+\Delta t$ and $t=\infty$. This results in:
\begin{eqnarray}
&&\hspace{-6mm}\phi_b\! -\!\phi_0=(1\!+\!\beta^{-1})(\theta_b\!-\!\theta_0)\!-\!k_0^{-1}\left[\Psi'(\theta_b)\!-\!\Psi'(\theta_0)\right],\nonumber\\
&&\hspace{-6mm}\phi_a \!-\!\phi_b= (1\!+\!(\beta\xi)^{-1})(\theta_a\!-\!\theta_b)\!-\!k_0^{-1}\left[\Psi'(\theta_a)\!-\!\Psi'(\theta_b)\right]\!\!,\\
&&\hspace{-6mm}\phi_\infty\! -\!\phi_a=(1\!+\!\beta^{-1})(\theta_\infty\!-\!\theta_a)\!-\!k_0^{-1}\left[\Psi'(\theta_\infty)\!-\!\Psi'(\theta_a)\right],\nonumber
\end{eqnarray}
where $\theta_0\!=\!\theta(t\!=\!0)\!\simeq\! 1$, $\theta_b=\theta(t_0)$ and $\theta_a=\theta(t_0\!+\!\Delta t)$ are the values of $\theta$ right before and right after the quarantine, respectively, and $\theta_{\infty}=\theta(t\to\infty)$ (and the same notation applies to $\phi$). Summing up these equations, and noting that $\dot{\theta}_0\simeq\dot{\theta}_{\infty}\simeq 0$ and thus, $\phi_0\simeq\phi_{\infty}\simeq 0$, and that $\Psi'(\theta_0)\simeq\Psi'(1)=k_0$, we arrive at an equation for $\theta_\infty$:
\begin{equation}
\label{eqn:thetaInf}
   \hspace{-2mm} \left(1\!+\!\frac{1}{\beta}\right)\left(\theta_{\infty}-1\right)+\frac{\Delta \theta}{\beta}\left(1\!-\!\frac{1}{\xi}\right)-\frac{\Psi'(\theta_{\infty})}{k_0}+1 = 0,
\end{equation}
where $\Delta \theta(\Delta t,S_b) \equiv \theta_b - \theta_a$. To find $S_{\text{opt}}$ we differentiate Eq.~(\ref{eqn:thetaInf}) with respect to $S_b$ which yields:
\begin{equation}
\label{eqn:optimalCondNet}
\frac{d\theta_\infty}{dS_b} = \frac{k_0\left(1-\xi\right)}{\xi\left[k_0+k_0 \beta -\beta \Psi''(\theta_{\infty})\right]}\frac{d\Delta\theta}{d\theta_b}.
\end{equation}
Thus, to minimize $R_\infty$ (maximize $\theta_\infty$), we equate $\Delta\theta'(\theta_b) \!=\! 0$ and find $\theta_{\text{opt}}$---the optimal value of $\theta_b$. This allows finding $S_{\text{opt}}$ [Eq.~(\ref{eqn:St})], which  minimizes $R_{\infty}$. 

Note that, in order to find the optimal susceptible fraction at the onset of quarantine, it is sufficient to numerically solve the rate equation during the quarantine. Indeed, starting from Eq.~(\ref{eqn:milleq}) and differentiating the first of Eqs.~(\ref{eqn:dynamicsMiller}) with respect to time (during the quarantine), we arrive at a second-order differential equation for $\theta$. Integrating this equation between $t=t_0$ and $t=t_0+\Delta t$, and noting that $\theta(t)$ is continuous at all times, we arrive at the following first-order equation for $\theta$:
\begin{equation}
    \label{eqn:phidur}
    \dot{\theta} = -\left(1+\beta \xi\right) \theta(t) +\xi+\theta_b(1-\xi)+(\beta \xi/k_0)\Psi'[\theta(t)],
\end{equation}
which is numerically solved with the initial condition $\theta(t_0)=\theta_b$. 
The solution of this equation allows finding $\Delta\theta$ as function of $\theta_b$, where the latter is arbitrary. Differentiating $\Delta\theta$ with respect to $\theta_b$ and equating to zero yields $\theta_{\text{opt}}$ and the optimal susceptible fraction.

While we have outlined a complete recipe of finding the optimal fraction of susceptibles that minimizes $R_\infty$, explicit analytical progress can be made in the limit of short quarantine, $\Delta t\ll 1$, as was done in the well-mixed topology. To do so, we take Eq.~(\ref{eqn:phidur}), estimate it at time $t=t_0 + \Delta t/2$, and use the approximations $\dot{\theta} \simeq -\Delta \theta/\Delta t$ and $\theta(t) \approx (\theta_b+\theta_a)/2 = \theta_b-\Delta \theta/2$. This yields an approximate equation for $\Delta \theta$, valid at $\Delta t\ll 1$:
\begin{equation}
\label{eqn:approxSmallDtDurCatas}
     \frac{\Delta \theta}{\Delta t} \simeq \theta_b\xi(1+\beta) -\frac{\Delta \theta}{2}\left(1+\beta \xi\right)-\xi-\frac{\beta \xi}{k_0} \Psi'\left(\!\theta_b-\frac{\Delta \theta}{2}\!\right)\!.
\end{equation}
Taylor-expanding $\Psi'(\theta_b-\Delta \theta/2) \approx \Psi'(\theta_b)-\Psi''(\theta_b) \Delta \theta/2$, we can thus write $\Delta \theta$ in terms of $\Delta t$ (up to second order):
\begin{eqnarray}
        \hspace{-5mm}\Delta \theta=\Delta \theta(\theta_b) &\simeq& \left[\theta_b(\beta+1)-1-\frac{\beta \Psi'(\theta_b)}{k_0}\right]\xi \Delta t \nonumber\\
        &\times& \left[1-\left(1+\beta \xi-\frac{\beta \xi \Psi''(\theta_b)}{k_0}\right)\frac{\Delta t}{2}\right].
\end{eqnarray}
To find $\theta_{\text{opt}}$ [see Eq.~(\ref{eqn:optimalCondNet})], we need to differentiate this equation with respect to $\theta_b$ and equate to zero, yielding:
\begin{eqnarray}
\label{eqn:DDeltaTheta}
    &&\hspace{-4mm}0=\frac{d\Delta \theta}{d\theta_b} =  \left[\left(\beta+1\right) k_0-\beta \Psi''(\theta_b)\right]\frac{\xi \Delta t}{k_0} \\ 
    &&\hspace{-4mm}+\left\{\beta\xi\left[k_0\left(-1+\theta_b+\beta \theta_b\right)-\beta \Psi'(\theta_b)\right]\Psi'''(\theta_b) \right. \nonumber\\
    &&\hspace{-4mm}-\left.\left[\left((\beta\!+\!1)k_0 \!-\!\beta \Psi''(\theta_b)\right)\left(k_0 \!+\!k_0\beta\xi\!-\!\beta\xi\Psi''(\theta_b)\right)\right]\right\}\frac{\xi \Delta t^2}{2k_0^2}.\nonumber
\end{eqnarray}
We now solve this equation perturbatively, by plugging $\theta_b=\theta_{\text{opt}}\!\simeq\!\theta_{\text{opt}}^{(0)}\!+\!\theta_{\text{opt}}^{(1)}\Delta t$ into~(\ref{eqn:DDeltaTheta}). In the leading-order in $\Delta t\ll 1$, we arrive at an implicit equation for $\theta_{\text{opt}}^{(0)}$:
\begin{equation}
\label{eqn:zerothOrderThetaC}
    \Psi''\left(\theta_{\text{opt}}^{(0)}\right) = (\beta+1)k_0/\beta.
\end{equation}
The next-order correction, $\theta_{\text{opt}}^{(1)}$, is found by plugging  $\theta_{\text{opt}}^{(0)}$ in the ${\cal O}(\Delta t^2)$ term in Eq.~(\ref{eqn:DDeltaTheta}). This yields:
\begin{equation}
   \label{eqn:FirstOrderThetaC}\theta_{\text{opt}}^{(1)} = \frac{\xi}{2}\left[-1+(1+\beta)\theta_{\text{opt}}^{(0)}-(\beta/k_0) \Psi'(\theta_{\text{opt}}^{(0)})\right].
\end{equation}
Having found $\theta_{\text{opt}}\simeq\theta_{\text{opt}}^{(0)}+\theta_{\text{opt}}^{(1)}\Delta t$,  the optimal fraction of susceptibles at the onset of quarantine reads 
\begin{equation}\label{eqn:general_Sc}
    S_{\text{opt}}\simeq \Psi\left(\theta_{\text{opt}}^{(0)}\right)+\Psi'\left(\theta_{\text{opt}}^{(0)}\right)\theta_{\text{opt}}^{(1)}\Delta t.
\end{equation}
We will now apply these results to three examples of networks: homogeneous, Poisson, and gamma networks.

\subsubsection{Homogeneous Network}
In a homogeneous network each node has an equal number of $k_0$ neighbors, and the degree distribution satisfies $p_k = \delta_{k,k_0}$. Thus $\Psi(\theta)$ [Eq.~(\ref{eqn:St})] becomes
\begin{equation}
    \Psi(\theta) = \theta^{k_0},
\end{equation}
or alternatively, $\theta = S^{1/k_0}$. As a result, in the general case one can solve Eq.~(\ref{eqn:phidur}) and find the optimal fraction of susceptibles $S_{\text{opt}}$, by numerically differentiating $\Delta \theta (\Delta t,S_b) =S_b^{1/k_0}-S_a^{1/k_0}$ with respect to $\theta_b$.

In the limit of short quarantine times, $\Delta t\ll 1$, we use Eqs.~(\ref{eqn:zerothOrderThetaC}) and (\ref{eqn:FirstOrderThetaC}) to find $\theta_{\text{opt}}^{(0)}$ and $\theta_{\text{opt}}^{(1)}$. Substituting these expressions into Eq.~(\ref{eqn:general_Sc}) and taking the limit of $k_0\gg 1$, we arrive at:
\begin{equation}
\label{ScOptSmallDtHom}
      S_{\text{opt}} \simeq \frac{1}{R_0}\left(1\!+\!\frac{1\!+\!R_0\!-\!2\ln{R_0}}{k_0} \right)+ \left( \frac{R_0 \!-\! \ln{R_0}\!-\!1}{2 R_0} \right) \xi \Delta t,  
\end{equation}
where we have omitted ${\cal O}(k_0^{-1})$ terms in the ${\cal O}(\Delta t)$ term.
Note that, in the limit $k_0 \to \infty $ we restore the result for the optimal fraction of susceptibles at the onset of quarantine, for a well-mixed network.

\subsubsection{Poisson Network}
In a Poisson network (also known as the Erd\"os-R\'enyi network~\cite{erdos1959random} for large networks), the degree distribution follows a Poisson distribution $p_k = (1/k!)k_0^k e^{-k_0}$, which has a mean degree of $k_0$. Here, $ \Psi(\theta)$ [Eq.~(\ref{eqn:St})] becomes
\begin{equation}
    \Psi(\theta) = e^{-k_0}(e^{k_0 \theta}-1),
\end{equation}
or alternatively, $\theta = (1/k_0)\ln (e^{k_0}S+1)$. As a result, in the general case one can solve Eq.~(\ref{eqn:phidur}) and find $S_{\text{opt}}$ by numerically differentiating $\Delta \theta (\Delta t,S_b) =(1/k_0)\ln\left[(S_b + e^{-k_0})/(S_a + e^{-k_0})\right]$ with respect to $\theta_b$.

In the limit of short quarantine times, $\Delta t\ll 1$, we use Eqs.~(\ref{eqn:zerothOrderThetaC}) and (\ref{eqn:FirstOrderThetaC}) to find $\theta_{\text{opt}}^{(0)}$ and $\theta_{\text{opt}}^{(1)}$. Substituting these  into Eq.~(\ref{eqn:general_Sc}) and taking the limit of $k_0\gg 1$, yields:
\begin{equation}
    \label{ScOptSmallDtPoiss}
        S_{\text{opt}} \simeq \frac{1}{R_0} \left(1+\frac{1+R_0}{k_0}\right)+ \left( \frac{R_0\!-\!\ln{R_0}\!-\!1}{2 R_0}\right) \xi\Delta t,
\end{equation}
where we have neglected the term $e^{-k_0}$ in the zeroth-order in $\Delta t$, as this term is exponentially small when $k_0\gg 1$. In addition, as done in the homogeneous network case, we have omitted ${\cal O}(k_0^{-1})$ terms in the ${\cal O}(\Delta t)$ term. The fact that the Poisson distribution has a relative width scaling as $k_0^{-1/2}\ll 1$,  gives rise to the resemblance to the homogeneous case. Finally, in the limit $k_0 \to \infty $, we again restore the well-mixed result for $S_{\text{opt}}$.

%Note that, the resemblance to the homogeneous case  stems from the fact that for $k_0\gg 1$, the Poisson distribution becomes very narrow with a relative width scaling as $k_0^{-1/2}$. Finally, in the limit $k_0 \to \infty $, we again restore the result for $S_{\text{opt}}$, for a well-mixed network.

\subsubsection{Gamma distribution}
We now consider a more general class of networks with two parameters such that the mean and standard deviation can be controlled separately. As a prototypical example we consider the gamma distribution, which is used in this realm to describe population heterogeneity~\cite{julicher2020,leibenzon2024heterogeneity}. In this case, the degree distribution satisfies
\begin{equation}
   p_k = \frac{k^{1/\epsilon^2\!-\!1}\, e^{-k/(k_0 \epsilon^2)}}{\Gamma(1/\epsilon^2) (k_0 \epsilon^2)^{1/\epsilon^2}}.
\end{equation}
Here, we have adjusted the shape and scale parameters such that the mean and variance respectively are $k_0$ and $\sigma^2=\epsilon^2 k_0^2$, while $\epsilon=\sigma/k_0$ is the distribution's coefficient of variation.
As a result, $ \Psi(\theta)$ [Eq.~(\ref{eqn:St})] becomes
\begin{equation}
        \Psi(\theta) = \left[1-k_0 \epsilon^2 \ln(\theta)\right]^{-1/\epsilon^2},
\end{equation}    
or alternatively, $\theta =\exp[(1-S^{-\epsilon^2})/(k_0\epsilon^2)]$. As a result, in the general case one can solve Eq.~(\ref{eqn:phidur}) and find the optimal fraction of susceptibles by numerically differentiating $\Delta \theta (\Delta t,S_b)$ with respect to $\theta_b$.
 
Unfortunately, the limit of short quarantine duration $\Delta t\ll 1$, is not amenable in the general case here, as Eq.~(\ref{eqn:zerothOrderThetaC}) cannot be solved analytically. Yet, we can solve the short-duration case numerically, see below. Nevertheless, analytical progress can be made in the limit of small coefficient of variation, $\epsilon \ll 1$, or $\sigma\ll k_0$, see below.

\subsubsection{General network}
Here we formulate an expression for the optimal fraction of susceptibles $S_{\text{opt}}$, which minimizes the expected outbreak size. We do so for generic networks, with an arbitrary degree distribution $p_k$, but under three assumptions: (i) the distribution's mean is large $k_0\gg 1$, (ii) the coefficient of variation is small, $\epsilon\ll 1$, and (iii) the duration of the quarantine is short, $\Delta t\ll 1$. 

Under these assumptions we begin by computing the subleading ${\cal O}(\Delta t)$ correction to $S_{\text{opt}}$, see Eqs.~(\ref{eqn:FirstOrderThetaC}) and~(\ref{eqn:general_Sc}). To do so, we use a very crude approximation, assuming that the distribution is a delta function peaked at $k=k_0$, namely $p_k = \delta_{k,k_0}$, as in the homogeneous case. As a result, we have $\Psi(\theta)\sim \theta^{k_0}$. Plugging this result into Eq.~(\ref{eqn:zerothOrderThetaC}) and putting $\beta\simeq R_0/k_0$ (valid when $\sigma\ll k_0$), we find $\theta_{\text{opt}}^{(0)}\simeq R_0^{-1}(1+\alpha/k_0)$, where $\alpha\ll k_0$ is a constant depending on $R_0$ and the higher central moments of the degree distribution. Consequently, we plug this result into Eq.~(\ref{eqn:FirstOrderThetaC}) and 
use the fact that $
\Psi'(\theta) \sim k_0 \theta^{k_0-1}$. In the leading order in $k_0\gg 1$, this yields
$\theta_{\text{opt}}^{(1)}\simeq \xi\left(R_0 - \ln{R_0} - 1\right)/(2k_0)$. As a result, the optimal susceptible fraction [Eq.~(\ref{eqn:general_Sc})] becomes:
\begin{equation}
\label{eqn:proofGeneralForm}
S_{\text{opt}} \approx \Psi\left(\theta_{\text{opt}}^{(0)}\right) + \left( \frac{R_0 - \ln{R_0} - 1}{2 R_0} \right) \xi \Delta t,
\end{equation}
where the leading-order term given by Eq.~(\ref{eqn:St}) is not yet calculated. It is implicitly given by Eq.~(\ref{eqn:zerothOrderThetaC}) and we now show how this term can also be approximately found by using a slightly more subtle approximation.

To compute the leading-order term, it is not sufficient to assume that the distribution is a delta function around $k_0\gg 1$. Rather, we assume that it is a narrow distribution with a standard deviation $\sigma\ll k_0$. For such networks, $\Psi(\theta)$ can in principle be evaluated using a saddle point approximation. Indeed, in this regime we can approximate
$\Psi(\theta)$ given by Eq.~(\ref{eqn:St}) as an integral and evaluate it via the saddle point method:
\begin{equation}
    \Psi(\theta)\simeq \int_0^{\infty} p(k) \theta^k dk\simeq A\, p(k_0)\theta^{k_0},
\end{equation}
where we have assumed that the saddle point $k^*$ approximately equals $k_0$, and $A$ is a prefactor depending on the width of the integrand in the vicinity of $k^*$. As a result, the second derivative of $\Psi$ satisfies
\begin{equation}\label{psipp}
    \hspace{-3mm}\Psi''(\theta)\simeq\int_0^{\infty} \frac{k(k-1)}{\theta^2} p(k) \theta^k dk\simeq  \frac{k_0(k_0-1)}{\theta^2}\Psi(\theta),
\end{equation}
where the slowly-varying prefactors were taken out of the integral and evaluated at $k^*\simeq k_0$. 
This allows solving Eq.~(\ref{eqn:zerothOrderThetaC}) for the leading-order of $S_{\text{opt}}$, by plugging Eq.~(\ref{psipp}) into~(\ref{eqn:zerothOrderThetaC}). Here, we still have an unknown prefactor of $\theta^{-2}$, but it turns out that for the sake of its calculation, it suffices to take $p_k=\delta_{k,k_0}$ which yields $\theta \sim \Psi^{1/k_0}$. Hence, in the leading order in $k_0\gg 1$ we find
\begin{equation}\label{S0gen}
\Psi\left(\theta_{\text{opt}}^{(0)}\right)\simeq \frac{1}{R_0}\left(1+\epsilon^2+\frac{1+R_0-2\ln{R_0}}{k_0} \right).
\end{equation}
Here, we have kept ${\cal O}(\epsilon^2)$ terms  but neglected ${\cal O}(\epsilon^2/k_0)$ terms.
This result is completely generic for any distribution with a mean degree of $k_0$ and coefficient of variation $\epsilon$, as long as $\epsilon\ll 1$. Plugging Eq.~(\ref{S0gen}) into~(\ref{eqn:proofGeneralForm}) we have found the optimal fraction of susceptibles at the onset of quarantine which minimizes $R_{\infty}$ for generic networks.

Notably, while the estimation of the HDI threshold (i.e., solving $\dot{I}=0$) can only be done numerically in the case of generic heterogeneous networks, Eq.~(\ref{S0gen}) provides an approximate analytic expression for this threshold for generic networks with a narrow degree distribution. We show this numerically in the following section.

\section{Results\label{sec:results}} 
 We now present results for $S_{\text{opt}}$---the optimal fraction of susceptibles at the onset of quarantine that minimizes $R_\infty$. We do so for the various networks we have studied, and check our theoretical predictions for $S_{\text{opt}}$~(\ref{eqn:general_Sc}) by using two numerical methods: (i) solving the deterministic rate equations and (ii) performing network-based simulations. Notably, through this validation of the theoretical results, we aim to gain a deeper understanding of the optimal quarantine strategies across different networks.

The first numerical method includes solving the deterministic rate equations, in which at time $t=t_0$, a temporary quarantine is initiated for a duration of $\Delta t$. We do so over a wide range of \( S_b \) values and determine the final outbreak fraction \( R_{\infty} \) for each case. By fitting a parabolic curve to the function \( R_{\infty}(S_b) \), we identify the optimal \( S_b \) that minimizes \( R_{\infty} \).

The second numerical method includes performing network-based simulations. For this, we have constructed networks that follow the prescribed degree distributions using the configurational model, which gives rise to networks with negligible degree-degree correlations~\cite{fosdick2018configuring}. In this work we have focused on four different types of networks: well-mixed, homogeneous, Poisson and gamma networks. Nevertheless, the formalism is generic and we can simulate any arbitrary network (including empirical networks) with a given degree distribution. For each network, we have used the Gillespie algorithm~\cite{gillespie1976general,gillespie1977exact,gillespie2007stochastic} and ran Monte Carlo (MC) simulations for each \( S_b \) value. As done with the rate equations, we identified the optimal \( S_b \) by determining which value minimizes \( R_{\infty} \).

\begin{figure}[t!]  
    \centering
    \includegraphics[width=0.9\linewidth]{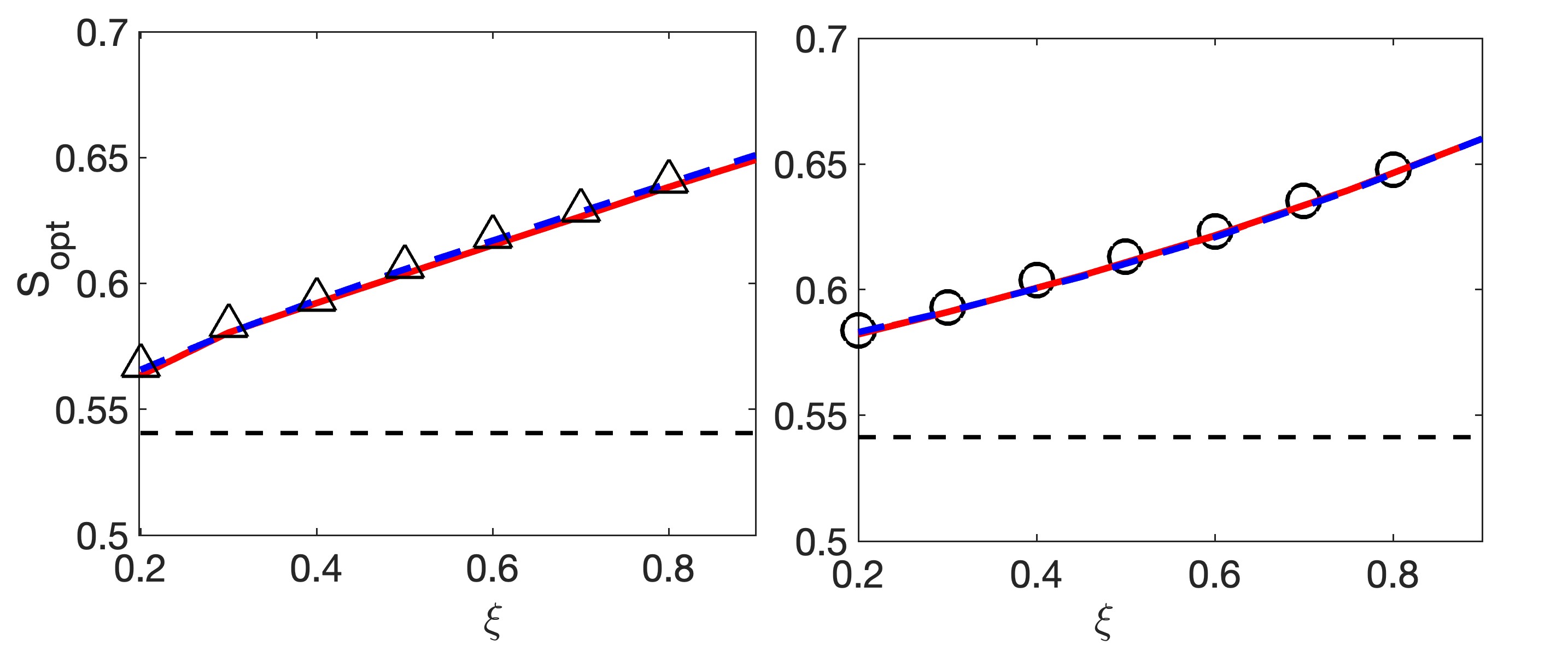} 
    \vspace{-4mm}
    \caption{Optimal fraction of susceptibles, $S_{\text{opt}}$, versus $\xi$ for $R_0 = 1.85$ and quarantine duration $\Delta t = 2$. Left and right panels show results for a well-mixed and Poisson network, respectively, with $k_0 = 80$. In both panels a numerical solution of the rate equations (solid lines) is compared to the theoretical prediction [Eqs.~(\ref{eqn:ScOptSmallDtWell}) and (\ref{ScOptSmallDtPoiss})] (dashed lines) and MC simulations (triangles and circles).
    Horizontal dashed line marks the HDI threshold in the absence of a quarantine.}
        \label{fig3}
\end{figure}

We have already established the existence of an optimal $S_b$ in Fig.~\ref{fig2} for two different networks, by showing heatmaps of $R_{\infty}$ versus $S_b$, while varying both $R_0$ and $\Delta t$. In Fig.~\ref{fig3}, we compare our theoretical predictions for $S_{\text{opt}}$ with a numerical solution of the rate equations and MC simulations, for well-mixed and Poisson networks. Here we show the dependence of $S_{\text{opt}}$ on the quarantine's strength, $\xi$. The agreement between the theoretical and numerical results is excellent. The dashed horizontal line indicates the HDI threshold which is approached as $\xi\to 0$. The HDI threshold equals $1/R_0$ for a well-mixed topology, and is calculated numerically for the Poisson network. Our results indicate that as the system approaches the HDI threshold, the number of infected individuals should increase at the onset of an optimal quarantine. Consequently, more stringent quarantine measures (lower $\xi$) are required to effectively minimize the overall expected outbreak size.

\begin{figure}[t!]  
    \centering
    \includegraphics[width=0.8\linewidth]{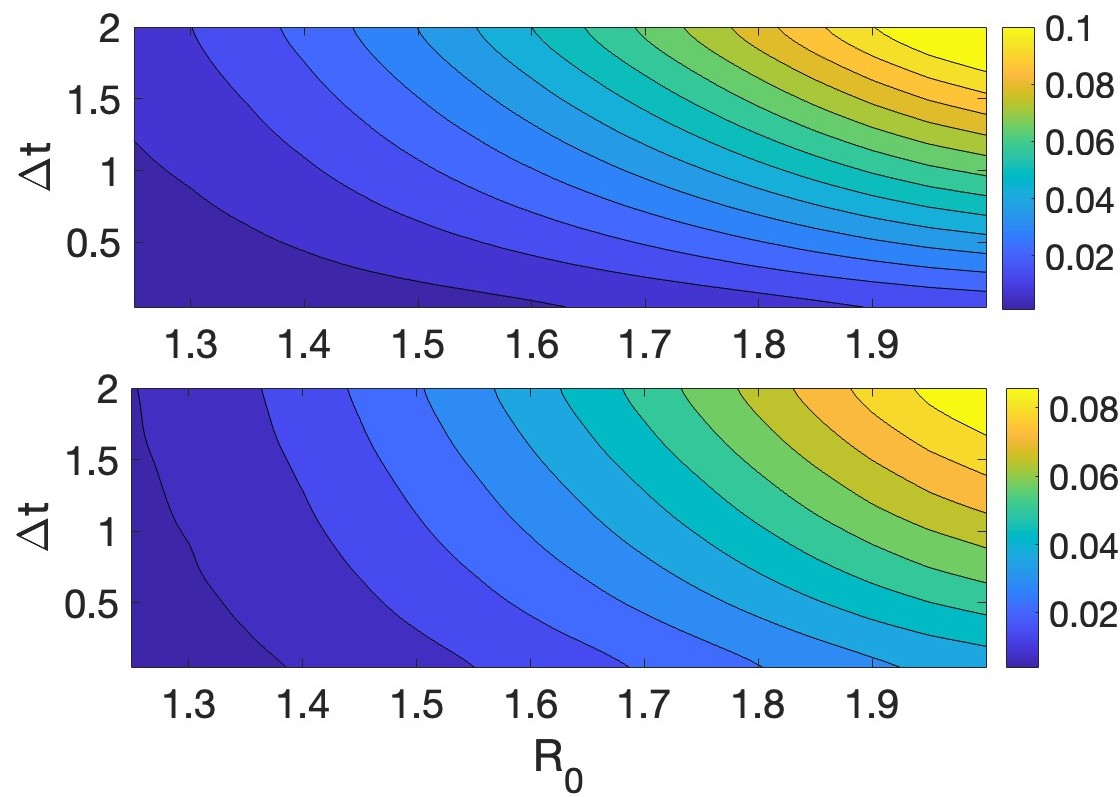} 
    \vspace{-5mm}
    \caption{Heatmaps of the relative distance between $S_{\text{opt}}$ and the HDI threshold, for a Poisson (upper panel) and gamma  (lower panel) network, versus $R_0$ and $\Delta t$. Warmer (cooler) colors indicate larger (smaller) relative distances. Here $k_0 = 20$, $\xi = 0.5$, and $\epsilon = 0.45$ (in the lower panel).}
        \label{fig4}
\end{figure}

In Fig.~\ref{fig4} we wanted to check our conjecture that as the quarantine duration decreases, $S_{\text{opt}}$ approached the HDI threshold  $S_{\text{hdi}}$ in the absence of quarantine. To this end, we plotted the relative difference $(S_{\text{opt}}-S_{\text{hdi}})/S_{\text{opt}}$ as a function of $R_0$ and  quarantine duration $\Delta t$, for Poisson and gamma networks. Indeed, as $\Delta t\to 0$, the relative distance vanishes. Also, as expected, the relative distance changes at a steeper rate as $R_0$ is increased, see Fig.~\ref{fig4}.

\begin{figure}[t!]  
    \centering \includegraphics[width=0.8\linewidth]{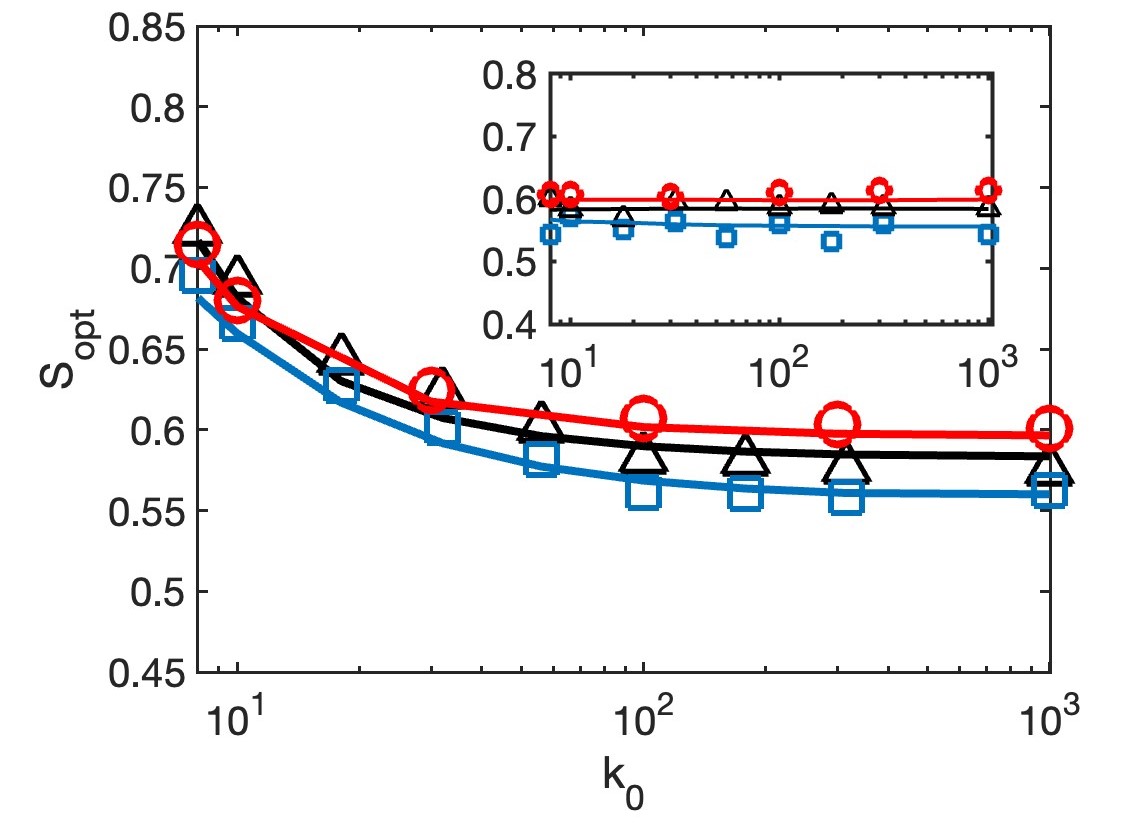} 
    \vspace{-5mm}
    \caption{
    Shown is $S_{\text{opt}}$ versus $k_0$ for $R_0 = 1.85$ , $\xi = 0.4$ and quarantine duration $\Delta t = 2$. Here we compare the numerical solution of the rate equations (solid lines) and MC simulation (symbols), for three different networks: homogeneous (triangles), Poisson (squares), and gamma (circles). For the latter we took a coefficient of variation of $\epsilon = 0.25$.
   Inset shows the same comparison, but here $R_0$ varies with $k_0$ such that $R_\infty$ in the absence of quarantine remains unchanged.}
    \label{fig5}
\end{figure}

In Fig.~\ref{fig5} we have studied the dependence of $S_{\text{opt}}$ on the average degree $k_0$, in the case of homogeneous, Poisson, and gamma networks. The figure shows that $S_{\text{opt}}$ increases as the mean degree decreases. Importantly, this indicates that as the network is more connected (with an increasing number of edges), the quarantine measures should be initiated at a later stage. Naturally, the reason is that by varying $k_0$, we effectively vary the critical infection rate below which the infection wave dies out instantly. As $k_0$ is decreased, $\beta_c$ is increased, see Eq.~(\ref{eqn:R0}), which means that by keeping $R_0$ constant, we effectively increase the infection rate. This means that the quarantine should start at an earlier stage, with less infected individuals. In the inset, we plot $S_{\text{opt}}$ as function of $k_0$ but vary $R_0$ in such a way that $R_\infty$ in the absence of quarantine remains the same. In this case, as seen in the inset $S_{\text{opt}}$ remains almost constant, as the effect of increasing $k_0$ is balanced by decreasing $R_0$. 

\begin{figure}[ht]
    \centering
    \includegraphics[width=0.8\linewidth]{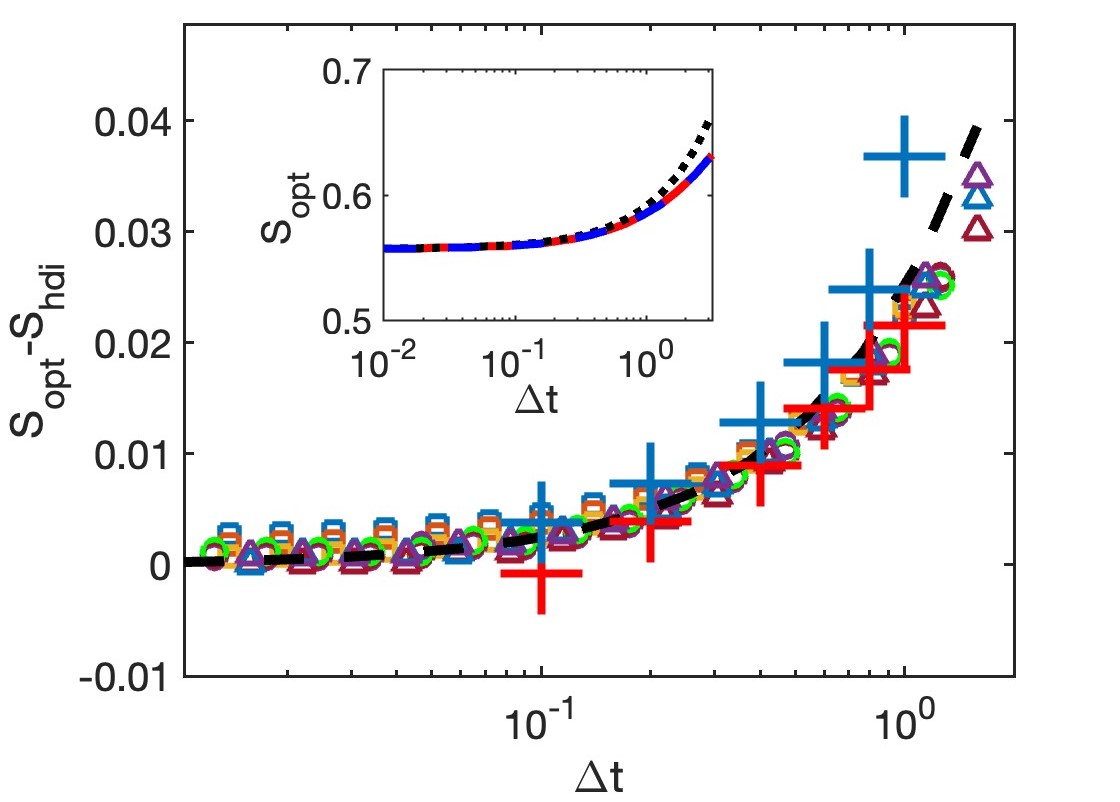}
    \vspace{-4mm}
    \caption{The difference $S_{\text{opt}}-S_{\text{hdi}}$ versus $\Delta t$, computed by numercially solving the rate equations, for homogeneous (triangles) and Poisson (squares) networks with $k_0 = 50$, $100$, $1000$, and for gamma networks (circles) with $k_0 = 50$ and $\epsilon = 0.01$, $0.05$, $1$. For all networks $R_0 = 1.7$. Plus symbols are MC simulations for a Poisson  (blue) and gamma  (red) network, with $k_0=80$ and $\epsilon = 0.25$ for the latter.
    Dashed line represents the theoretical prediction~(\ref{eqn:proofGeneralForm}). Inset shows $S_{\text{opt}}$ versus  $\Delta t$,  comparing between the numerical solution of the rate equations (solid line), semi-analytical prediction (dashed line) and approximated theoretical prediction (dotted line), in a  Poisson network with $k_0 = 50$ , $\xi = 0.5$, and  $R_0 = 1.9$.
    }
    \label{fig6}
\end{figure}

In Fig.~\ref{fig6} we further probe our conjecture that $S_{\text{opt}}$ coincides with the HDI threshold as the quarantine duration $\Delta t$ goes to zero. To this end, we plot the difference between $S_{\text{opt}}$ and $S_{\text{hdi}}$ as function of $\Delta t$, for a large variety of networks with varying $k_0$ and $\epsilon$. 

Figure~\ref{fig6} shows two important features. First, it proves that as $\Delta t\to 0$, $S_{\text{opt}}$ approaches $S_{\text{hdi}}$, and that $S_{\text{hdi}}\simeq \Psi(\theta_{\text{opt}}^{(0)})$, see Eqs.~(\ref{eqn:proofGeneralForm}) and (\ref{S0gen}). Although the HDI threshold cannot be computed analytically for arbitrary networks, our results strongly support the fact that the analytical expression in Eq.~(\ref{S0gen}) is an excellent approximation for $S_{\text{hdi}}$.  Second, the figure shows that the universal first-order correction in $\Delta t\ll 1$ that we have found for any arbitrary network, excellently agrees with numerical solutions of the deterministic rate equations and numerical MC simulations for a wide variety of network topologies. In the inset we show a comparison between the approximated theoretical prediction [Eq.~(\ref{ScOptSmallDtPoiss})], valid for short $\Delta t$ and large $k_0$, the semi-analytical prediction, found by numerically solving  Eq.~(\ref{eqn:phidur}) during the quarantine for $\Delta \theta(S_b)$, and the full numerical solution of the rate equations. One can see that the theoretical approximation remains  accurate as long as  $\Delta t\ll 1$.

%\vspace{0.7cm}
\section{Discussion and Conclusion \label{sec:discussion}}
%\vspace{-0.2cm}
We have studied the optimal timing for initiating temporary quarantine measures within the SIR model on heterogeneous  networks, in order to reduce epidemic impact.  By analyzing how the quarantine timing impacts the epidemic curve, we were able to determine the initiation point that would most effectively limit the spread, i.e., the fraction of susceptibles at the start of quarantine.

Our theoretical framework was based on the mean-field, deterministic theory, and we computed the optimal susceptible fraction $S_{\text{opt}}$ at the onset of the quarantine, which minimizes the final outbreak fraction, $R_\infty$. Our results were obtained for four prototypical network structures: well-mixed (fully connected), homogeneous, Poisson and gamma networks. Importantly, when the quarantine duration is not too long, we have derived analytical expressions for the optimal timing of the quarantine, also for generic networks with an arbitrary (narrow) degree distribution. Our results reveal that in this case, the optimal moment to begin quarantine occurs just before the herd immunity (HDI) threshold is reached. This finding is counterintuitive, as na\"ively one would think early intervention is  more efficient. However, initiating quarantine too early may only delay the epidemic’s peak rather than reduce its size, as it prevents sufficient immunity accumulation. 
We have also shown that for highly connected networks quarantine measures should be initiated at a later stage compared with sparse networks.

Our conclusions emphasize the importance of well-timed quarantine measures in effectively managing epidemics. Interventions aligned with the HDI threshold enable a balance between delaying the epidemic and promoting the development of natural immunity, ultimately minimizing the total expected outbreak size.

Future directions include incorporating stochastic effects such as demographic noise into the model to capture the variability and unpredictability of real-life epidemics with a finite population. Furthermore, exploring ways to determine the HDI threshold directly from timing the optimal quarantine may become a practical tool for epidemic control in networked populations, potentially aiding in more responsive public health decision-making.

\bibliography{bib}
\end{document}